\documentclass[aps,showpacs,nofootinbib,amsmath,amssymb,preprint]{revtex4}
\usepackage{bm}
\usepackage{graphicx}

\begin{document}

\title{
Robustness of the $S$-deformation method for black hole stability analysis
}

\author{
${}^{1}$Masashi Kimura
and
${}^{2,3}$Takahiro Tanaka,
}

\affiliation{
${}^{1}$CENTRA, Departamento de F\'{\i}sica, Instituto Superior T\'ecnico, Universidade de Lisboa, Avenida~Rovisco Pais 1, 1049 Lisboa, Portugal
\\
${}^{2}$Department of Physics, Kyoto University, Kyoto 606-8502, Japan
\\
${}^{3}$Center for Gravitational Physics, Yukawa Institute for Theoretical Physics, Kyoto University, 
Kyoto 606-8502, Japan
}

\date{\today}
\preprint{KUNS-2723}
\preprint{YITP-18-52}

\begin{abstract}
The $S$-deformation method is a useful way to show the linear mode stability of a black hole when the perturbed field equation takes the form of the Schr\"odinger equation. While previous works where many explicit examples are studied suggest that this method works well, general discussion is not given yet explicitly. In this paper, we show the existence of a regular $S$-deformation when a black hole spacetime is stable under some reasonable assumptions. This $S$-deformation is constructed from a solution of a differential equation. We also show that the boundary condition for the differential equation which corresponds to a regular $S$-deformation has a one-parameter degree of freedom with a finite range. This is the reason why any fine-tune technique is not needed to find $S$-deformation numerically.
\end{abstract}

\maketitle

\section{Introduction}

When the black hole spacetime is highly symmetric, the linear gravitational perturbation
equation usually takes the form of the two-dimensional wave equation~\cite{
Regge:1957td, Vishveshwara:1970cc, Zerilli:1970se, Zerilli:1971wd, 
Kodama:2003jz, Ishibashi:2003ap, Kodama:2003kk, 
Dotti:2004sh, Dotti:2005sq, Gleiser:2005ra, Takahashi:2010ye, Takahashi:2009dz, Takahashi:2009xh}
\begin{align}
\left[
-\frac{\partial^2 }{\partial t^2}  +\frac{\partial^2 }{\partial x^2}  - V(x) \right]\tilde{\Phi} = 0.
\label{mastereqtr}
\end{align}
Using the Fourier transformation with 
respect to the time coordinate, 
$\tilde{\Phi}(t,x) = e^{-i\omega t}\Phi(x)$, this equation 
takes the form of the Schr\"odinger equation 
\begin{align}
\left[
-\frac{d^2}{dx^2} + V\right]\Phi = \omega^2 \Phi =: E \Phi.
\label{schrodingereq}
\end{align}
It is known that the non-existence of a bound state with $E < 0$
implies the non-existence of an exponentially growing mode in time, 
{\it i.e.}, the linear mode stability of the black hole.

For a continuous function $S$, from Eq.\eqref{schrodingereq}, we can show
\begin{align}
-
\left[
\Phi^*\frac{d \Phi}{dx} + S |\Phi|^2
\right]_{- \infty}^{\infty}
+
\int dx
\left[\left|\frac{d\Phi}{dx} + S \Phi \right|^2 +
\left(V + \frac{dS}{dx} - S^2 \right) \left|\Phi \right|^2 \right]
= E \int dx\left|\Phi \right|^2,
\label{sdef1}
\end{align}
where $\Phi^*$ is the complex conjugate of $\Phi$. We consider the boundary condition such that 
the boundary term vanishes.
We can regard that the potential is deformed as
\begin{align}
\tilde{V} := V + \frac{dS}{dx} - S^2,
\end{align}
which is called the $S$-deformation.
If $\tilde{V}$ is 
non-negative everywhere by choosing an appropriate function $S$, we can say 
$E \ge 0$ for any $\Phi$, {\it i.e.,}
the non-existence of a negative energy bound state. This method was 
introduced in~\cite{Kodama:2003jz, Ishibashi:2003ap, Kodama:2003kk}
and used for the stability analysis in many cases, {\it e.g.} in~\cite{Dotti:2004sh, Dotti:2005sq, Gleiser:2005ra, 
Takahashi:2010ye, Takahashi:2009dz, Takahashi:2009xh, Beroiz:2007gp, Takahashi:2010gz}.

In the previous work~\cite{Kimura:2017uor}, 
a simple method to find a regular $S$ numerically 
by solving an equation $\tilde{V} = 0$, {\it i.e.,}
\begin{align}
V + \frac{dS}{dx} - S^2 = 0,
\label{sdefvzero}
\end{align}
was discussed,
when it is hard to find it analytically.
In some examples, the existence of a regular solution 
of Eq.~\eqref{sdefvzero} 
was used to show that
the spacetime is stable~\cite{Kimura:2017uor},
but it is not known whether a regular solution always exists or not when the spacetime is stable.
In fact, Eq.~\eqref{sdefvzero} is 
satisfied by
\begin{align}
S = -\frac{1}{\Phi_{E = 0}} \frac{d\Phi_{E = 0}}{dx},
\end{align}
with a solution 
of the Schr\"odinger equation with zero energy $\Phi_{E = 0}$.\footnote{
Note that this $\Phi_{E = 0}$ is not necessary square integrable, and usually
it corresponds to a growing mode.} 
Thus, 
the existence of $\Phi_{E = 0}$ which does not have zero 
anywhere 
implies the existence of a regular solution of Eq.~\eqref{sdefvzero}.
In this paper, we show the existence of such a wave function when the spacetime is stable (under some assumptions).
This result is an almost immediate consequence 
of the Sturm-Liouville theory,
but we think that
this fact is not well recognized in the community of the black hole stability
and
the proof does not seem trivial when the domain of the potential $V$ is $-\infty < x < \infty$.
We also discuss the robustness of this method, {\it i.e.,}
we can find a regular $S$-deformation numerically without fine-tuning.

\section{$S$-deformation in a finite box}
The discussion in this section is almost trivial from the 
usual Sturm-Liouville  theory, but we think that 
pedagogically it is still worth giving an explicit proof.
We focus on the domain where the potential is bounded 
and piecewise continuous.
First, we introduce a proposition ({\it e.g.}, see~\cite{MessiahvolI, Moriconi:2007}.)

\vspace{10pt}\hspace{-15pt}{\bf Proposition 1.}~~{\it 
Let us consider two solutions $\Phi_1$ and $\Phi_2$ of the Schr\"odinger equation~\eqref{schrodingereq} for energies 
$E_1$ and $E_2$ with $E_1 < E_2$, respectively.
If $\Phi_1$ has two consecutive zeros at $x = x_L$ and $x = x_R$, $\Phi_2$ has at least one zero 
between $x_L$ and $x_R$.
}
\\
{\bf Proof}.~~
From the Schr\"odinger equation~\eqref{schrodingereq} for energies $E_1$ and $E_2$, we have
\begin{align}
(\Phi_1^\prime \Phi_2 - \Phi_2^\prime \Phi_1)^\prime = (E_2 - E_1)\Phi_1 \Phi_2,
\label{eq:1}
\end{align}
where a prime denotes the derivative with respect to $x$.
We can assume $\Phi_1 > 0$ for $x_L < x <x_R$ without loss of generality.
Then, $\Phi_1^\prime|_{x_L} > 0$ and $\Phi_1^\prime|_{x_R} < 0$ are satisfied.\footnote{
If $\Phi_1^\prime|_{x_L}=0$ or $\Phi_1^\prime|_{x_R} = 0$, $\Phi_1$ becomes a trivial solution, {\it i.e.,}
$\Phi_1 = 0$ everywhere, since both $\Phi_1$ and its derivative are zero at $x = x_L$ or $x = x_R$.
However, this cannot happen because $\Phi_1 > 0$ for $x_L < x <x_R$.
}
Integrating this equation, we have
\begin{align}
\Phi_1^\prime|_{x_R} \Phi_2|_{x_R}
-
\Phi_1^\prime|_{x_L} \Phi_2|_{x_L} 
= (E_2 - E_1)\int_{x_L}^{x_R} dx \,\Phi_1 \Phi_2.
\label{integral1}
\end{align}
If $\Phi_2$ does not have zero in $x_L < x < x_R$, we have $\Phi_2 \ge 0$ for $x_L \le x \le x_R$.
In that case, while the left-hand side of Eq.\eqref{integral1} is not positive, the right-hand side of Eq.\eqref{integral1} is positive. This is a contradiction. Thus, 
$\Phi_2$ has a zero in $x_L < x < x_R$.\hfill$\Box$
\vspace{10pt}

{}From this proposition, we can immediately show the well known result: 
the $n$-th excited state has $n$ nodes. Also, we can show the following lemma.

\vspace{10pt}\hspace{-15pt}{\bf Lemma 1.}~~{\it 
If there exists a solution for 
$E = E_1$ which has two consecutive zeros at $x = x_L$ and $x = x_R$,
there also exists a solution for $E = E_0 (< E_1)$ which does not have zero in $x_L \le x \le x_R$.
}
\\
{\bf Proof}.~~
Solving the Schr\"odinger equation for $E_0$ from $x = x_L$ with the boundary condition 
$\Phi|_{x_L} = 0, \Phi|_{x_L}^\prime = 1$, we obtain a solution $\Phi_0$.
From the above proposition, $\Phi_0$ does not have zero for $x_L < x \le x_R$.
We denote $a = \Phi_0|_{x_R} (>0), b =\Phi_0^\prime|_{x_R}$.
Solving the Schr\"odinger equation for $E_0$ from $x = x_R$ with the boundary condition 
$\Phi|_{x_R} = a$ and $\Phi|_{x_R}^\prime = \tilde{b} (< b)$, we obtain a solution $\tilde{\Phi}_0$.
Integrating the Wronskian conservation 
equation $(\Phi_0^\prime \tilde{\Phi}_0 - \tilde{\Phi}_0^\prime \Phi_0)^\prime = 0$
from a point $y ~(x_R > y \ge x_L)$ to $x_R$, 
we have
\begin{align}
-
\Phi_0^\prime|_{y}  \tilde{\Phi}_0|_{y}  + \tilde{\Phi}_0^\prime|_{y}  \Phi_0|_{y}  = 
-a (b - \tilde{b}) (< 0).
\end{align}
If $\tilde{\Phi}_0|_{y} = 0$ 
and $\tilde{\Phi}_0$ is positive in $y < x < x_R$,
$\tilde{\Phi}_0^\prime|_y$ takes positive value.
Thus, the above equation implies that $\Phi_0|_x$ becomes negative.
However, this contradicts with that $\Phi_0 \ge 0$ for $x_L \le x < x_R$. Thus, 
$\tilde{\Phi}_0$ is positive for $x_L \le x \le x_R$
and this is the desired solution.
\hfill$\Box$

\vspace{10pt}
If the energy of the ground state is larger than zero, from Lemma 1,
we can construct a solution of the Schr\"odinger equation with zero energy which 
does not have zero anywhere when the potential is in a finite box, {\it i.e.}, 
$V$ is bounded
and piecewise continuous in $-L < x < L$ with a positive constant $L$, 
but $V = \infty$ in $|x| \ge L$.
Using such a solution $\Phi_{E=0}$, we can construct a regular $S$-deformation as
\begin{align}
S = - \frac{1}{\Phi_{E=0}}\frac{d\Phi_{E=0}}{dx}.
\end{align}
Thus, we have the following:

\vspace{10pt}\hspace{-15pt}{\bf Proposition 2.}~~{\it 
If the energy of the ground state is larger than zero and 
the potential is in a finite box, 
there exists a regular $S$-deformation.
}

\vspace{10pt}
Since this proposition holds for arbitrary value of $L$, 
we can expect that 
there also exists a regular $S$-deformation 
for a stable spacetime even when $L \to \infty$.
In the next section, we discuss the case with $L \to \infty$.

\section{infinite size box}
We consider the case in ``an infinite size box'', {\it i.e.},
the domain of our interest is $-\infty < x < \infty$.
Even in this case, if there exists a bound state with $E > 0$,
we can easily show that 
the non-existence of a bound state with negative energy implies
the existence of a regular $S$-deformation
by a similar discussion in Lemma 1.
In fact, this is a well known result ({\it e.g.,} see \cite{Panigrahi:1993zy}).
This condition is not usually satisfied in a black hole perturbation problem 
since $V \to 0$ at the horizon.
Hereafter, we do not assume the existence of a bound state with $E > 0$.

In this section, we assume the following conditions:
\\
(i) The potential $V$ is bounded and piecewise continuous in $-\infty < x < \infty$.
At $x \to \pm \infty$, $V$ takes non-negative constants $V_\pm$, respectively.
\\
(ii) For $E < 0$, the Schr\"odinger equation 
has solutions $\Phi^{L,\pm}_E(x)$ and $\Phi^{R,\pm}_E(x)$ 
whose leading behaviors 
are  $\Phi^{L,\pm}_E \simeq e^{\mp \sqrt{V_- - E} \, x + o(x)}$ at $x \to -\infty$
and $\Phi^{R,\pm}_E \simeq e^{\pm \sqrt{V_+ - E} \, x + o(x)}$ at $x \to \infty$, respectively.
\\
(iii) For $E = 0$, the Schr\"odinger equation  has 
solutions 
$\Phi^{L,-}_0(x)$ and $\Phi^{R,-}_0(x)$
which asymptote to a non-negative constant at $x \to -\infty$ and $x \to \infty$, respectively.

Hereafter, we omit to write $o(x)$
since this does not affects the following discussions.

\subsection{Existence of regular $S$ for stable spacetime}
First, we show the following three lemmas:

\vspace{10pt}\hspace{-15pt}{\bf Lemma 2.}~~{\it 
For an energy $E_1 < 0$, if 
there exists $x_1$ such that
$\Phi^{L,-}_{E_1}|_{x_1} = 0$ and $\Phi^{L,-}_{E_1} > 0$ for $-\infty < x < x_1$,
there exists a bound state with negative energy.
}
\\
{\bf Proof}.~~
For brevity, we denote
$\Phi^{L,-}_{E}$
as $\Phi_{E}$ in this proof.
Integrating the equation
$(\Phi_{E_1}^\prime \Phi_{E} - \Phi_{E}^\prime \Phi_{E_1})^\prime = (E - E_1)\Phi_{E_1} \Phi_E$
with $E < E_1$ from $-\infty$ to a point $y$,
we have
\begin{align}
(\Phi_{E_1}^\prime \Phi_{E} - \Phi_{E}^\prime \Phi_{E_1})|_y
= 
(E - E_1)\int_{-\infty}^{y} dx \,\Phi_{E_1} \Phi_E.
\end{align}
Let us assume $\Phi_E|_{x_2} = 0$ and $\Phi_E > 0$ for $-\infty < x < x_2$,
then $\Phi_E^\prime |_{x_2}$ is negative. 
If $x_2 \le x_1$, the above equation with $y = x_2$ becomes
\begin{align}
- \Phi_{E}^\prime|_{x_2} \Phi_{E_1}|_{x_2}
= 
(E - E_1)\int_{-\infty}^{x_2} dx\, \Phi_{E_1} \Phi_E.
\end{align}
Now the integrand is non-negative, and the integral is finite because 
$\Phi_E \simeq e^{\sqrt{V_- - E} x}$ at $x \to -\infty$.
Since the right hand side is negative and $\Phi_{E}^\prime|_{x_2}$ is negative,
$\Phi_{E_1}|_{x_2}$ should be negative. However, this contradicts with the fact that 
$\Phi_{E_1} \ge 0$ for $ -\infty < x \le x_2 \le x_1$.
Thus, $x_2$ should satisfies $x_1 < x_2$ if it exists.

If we consider $E \le V_{\rm min}$, where $V_{\rm min}$ is the minimum value of $V$, 
since $\Phi_E^{\prime \prime}/ \Phi_E = (V - E) \ge (V_{\rm min} - E) \ge 0$,
once $\Phi_E$ and $\Phi_E^\prime$ take positive value at some point, $\Phi_E$ remains to be positive as $x$ increases.
Since the boundary condition is $\Phi_E \simeq e^{\sqrt{V_- - E} x}$ at $x \to -\infty$,
we find $\Phi_E > 0$ everywhere for $E \le V_{\rm min}$.

Since the position of zero $x_2$ is a continuous function of $E (\le E_1)$ 
if it exists (see appendix~\ref{appendix:continuityzeroofphi})
and it can take all values in $x_1 < x_2 < \infty$ by changing $E$,\footnote{
Note that $\Phi_E$ is not necessary a continuous function of $E$, but we can say
the position of zero $x_2$ is a continuous function of $E$.
}
we can say that 
there exists $E_0~(V_{\rm min} < E_0 < E_1)$ 
such that $x_2 \to \infty$ for $E = E_0$.

The asymptotic behavior near $x \to \infty$ is
$\Phi_{E} \simeq c_1 e^{-\sqrt{V_+ - E}\, x} + c_2 e^{\sqrt{V_+ - E}\, x}$.
If $0 < 1 - E/E_0 \ll 1$, $x_2$ is in the asymptotic region.
So, from the condition $\Phi_{E}|_{x_2} = 0$, 
we obtain 
$\Phi_{E} \simeq c_1 (e^{-\sqrt{V_+ - E}\, x} - e^{\sqrt{V_+ - E}(x - 2 x_2)})$ when $0 < 1 - E/E_0 \ll 1$.
Thus, in the limit $E \to E_0 +0$, {\it i.e.,} $x_2 \to \infty$, the second term vanishes, 
and then $\Phi_{E_0} \simeq c_1 e^{-\sqrt{V_+ - E_0}\, x} \to 0$ at $x \to \infty$.
Thus, the solution $\Phi_{E_0}$ is the ground state.
\hfill$\Box$

\vspace{10pt}\hspace{-15pt}{\bf Lemma 3.}~~{\it 
If the Schr\"odinger equation for zero energy has 
a solution which has two 
consecutive zeros at $x = x_0$ and $x = x_1$,
there exists a bound state with negative energy.
}
\\
{\bf Proof}.~~
We denote the solution of 
the Schr\"odinger equation for zero energy 
which has two 
consecutive zeros at $x = x_0$ and $x = x_1$
as $\Phi_0$ in this proof.
Let us consider to solve the Schr\"odinger equation for $E \le 0$ with the 
boundary condition $\Phi|_{x_1} = 0$ and 
$\Phi^\prime|_{x_1} = \Phi_0^\prime|_{x_1}$. 
If we change the value of $E$ from zero, the position of zero also changes from $x = x_0$. 
From the similar discussion in Lemma 2, for some energy $E < 0$, the solution becomes decaying mode at $x \to -\infty$.
This implies the existence of a bound state with negative energy from Lemma 2.
\hfill$\Box$

\vspace{10pt}\hspace{-15pt}{\bf Lemma 4.}~~{\it 
If $S$-deformation function that satisfies Eq.\eqref{sdefvzero} 
is continuous for $-\infty < x < \ell$ with a constant $\ell$, $S^2$ asymptotes to 
$V_{-}$ at $x \to - \infty$.
}
\\
{\bf Proof}.~~
Since $S$ satisfies Eq.\eqref{sdefvzero} 
and the potential $V$ is bounded above and below, 
if $S$ is divergent in the asymptotic regions, 
the only possibility is that $dS/dx \simeq S^2$ and then $S \simeq 1/(c_2 - x)$.
However, it does not happen since $1/(c_2 - x)$ can be divergent only for finite $c_2$.
Thus, $S$ is also bounded above and below.

If $S$ oscillates and does not have a limit in $x \to -\infty$,
$S$ has infinite number of local maximum/minimum at $x = \alpha_n < \ell ~(n = 1, 2, \cdots)$ 
with $\alpha_{n+1} < \alpha_n$ and $\lim_{n\to \infty}\alpha_n \to -\infty$.
When $S$ is smooth, 
since $dS/dx = 0$ at each $x = \alpha_n$,\footnote{
$S$ can be non-smooth at $x = x_\alpha$ when $V$ is not continuous there.
In that case, $S$ is still continuous but
the two limits $\lim_{\epsilon \to 0}S^\prime|_{x = \alpha_n \pm \epsilon}$ with $\epsilon >0$ 
can be different values.
Since $x = \alpha_n$ is a local maximum/minimum of $S$,
$\lim_{\epsilon \to 0} (S^\prime|_{\alpha_n + \epsilon} S^\prime|_{\alpha_n - \epsilon})$
takes a negative value, 
and there exists positive constants $c_n, d_n$ such that 
$\lim_{\epsilon \to 0}(c_n S^\prime|_{\alpha_n + \epsilon} + d_n S^\prime|_{\alpha_n - \epsilon}) = 0$.
Thus, the equation
\begin{align}
S|_{\alpha_n}
&=
\lim_{\epsilon \to 0}
\frac{c_n V|_{\alpha_n + \epsilon} + d_n V|_{\alpha_n - \epsilon}}{c_n+d_n}
\end{align}
holds, and in the limit $n \to \infty$, the right hand side becomes $V_{-}$,
but the left hand side does not have a limit point. This is a contradiction,
and then $S$ should asymptote to a constant at $x \to -\infty$.
}we have
\begin{align}
V|_{\alpha_n} = S^2|_{\alpha_n}.
\end{align}
This implies that $V|_{\alpha_n} \neq V|_{\alpha_{n+1}}$ for an arbitrary large $n$, which
contradicts with the fact that $V$ asymptotes to $V_-$ at $x \to -\infty$.
So, $S$ should asymptote to a constant at $x \to -\infty$.

Since $S$ asymptotes to a constant at $x \to -\infty$,
from Eq.\eqref{sdefvzero}
\begin{align}
S = \int dx (- S^2 + V),
\end{align}
we can see that $S^2 \to V_{-}$ so that the integral is finite.
\hfill$\Box$
\\

We can show the existence of a regular $S$-deformation for a stable black hole as follows.

\vspace{10pt}\hspace{-15pt}{\bf Proposition 3.}~~{\it 
If bound states with negative energy do not exist, 
there exists a regular $S$-deformation function that satisfies Eq.\eqref{sdefvzero}.
}
\\
{\bf Proof}.~~
We assume that
$\Phi^{L,-}_0$ has zero at $x = x_1$.
If $\Phi^{L,-}_0$ has another zero except at $x \to \pm \infty$, 
there exists a bound state with negative energy from Lemma 3.
So, we only need to consider the case 
$\Phi^{L,-}_0 > 0$ in $-\infty < x < x_1$.
We denote $\Phi^{L,-}_0$ by $\phi_1$.
From Lemma 4, $(\phi_1^\prime / \phi_1)^2 \to V_-$ at $x \to -\infty$.
Since $\phi_1$ is not a growing mode at $x \to -\infty$, we can say
\begin{align}
\frac{\phi_1^\prime}{\phi_1} \to \sqrt{V_-},
\label{phi1vm}
\end{align}
at $x \to -\infty$.

Let us consider to solve the Schr\"odinger equation for zero energy
with the boundary condition $\Phi = 0$ at $x = x_2 > x_1$.
We denote the solution by $\phi_2$.
If $\phi_2$ has zero in $x_1 \le x < x_2$, 
this implies the existence of
a bound state with negative energy from Lemma 3.
So, we can consider $\phi_2 > 0$ in $x_1 \le x < x_2$.
From $(\phi_1 \phi^{\prime}_2-\phi^{\prime}_1 \phi_2)^\prime =0$,
we obtain a relation
\begin{align}
\phi_1 \phi^{\prime}_2-\phi^{\prime}_1 \phi_2 = \epsilon,
\label{epsilondef}
\end{align}
with a positive constant $\epsilon$ by evaluating Eq.\eqref{epsilondef} 
at $x = x_1$, where we used the relations 
$\phi_1|_{x_1} = 0, \phi^\prime_1|_{x_1} < 0, \phi_2|_{x_1} >0$.
Eq.\eqref{epsilondef} can be written in the form
\begin{align}
\frac{\phi^{\prime}_2}{\phi_2 }= \frac{\phi^{\prime}_1}{\phi_1} + \frac{\epsilon}{\phi_1\phi_2}.
\label{eqphi1phi2}
\end{align}
If $\phi_2 > 0$ in $-\infty < x< x_1$, 
from Eqs.\eqref{phi1vm},\eqref{eqphi1phi2} and Lemma 4,
we have 
\begin{align}
\frac{\epsilon}{\phi_1\phi_2} \to -\sqrt{V_-} \pm \sqrt{V_-} \le 0
\end{align}
at $x \to -\infty$. 
This is possible only when $\phi_2 \to \infty$ at $x \to -\infty$ since $\phi_1$ is not a growing mode.
However, setting $\phi_2=\phi_1 Z$, $Z$ satisfies
\begin{align}
Z^{\prime}= \frac{\epsilon}{\phi_1^2} \ge 0,
\end{align}
then $Z$ and $\phi_2$ cannot asymptotes to $+\infty$ at $x \to -\infty$.
Thus, $\phi_2$ should have a zero somewhere in $-\infty < x < x_1$.
This implies the existence of
a bound state with negative energy from Lemma 3.

Thus, 
$\phi_1$ does not have a zero except at $x \to -\infty$
if bound states with negative energy do not exist.
Then
\begin{align}
S = - \frac{1}{\phi_1}\frac{d\phi_1}{dx},
\end{align}
becomes a regular solution of Eq.\eqref{sdefvzero} 
in $-\infty < x < \infty$, 
since $S$ is continuous and it is bounded above and below from Lemma 4.
\hfill$\Box$

\subsection{Robustness of $S$-deformation method}
We can construct the general regular $S$-deformation $S$ that satisfies Eq.~\eqref{sdefvzero},
from $\Phi_0^{L,-}$ and $\Phi_0^{R,-}$, as follows: 

\vspace{10pt}\hspace{-15pt}{\bf Corollary 1.}~~{\it 
If there exists no bound state with $E \le 0$,
the general non-trivial solution of the Schr\"odinger equation for zero energy 
which does not have a node is given by 
$\Psi = c_L \Phi_0^{L,-} + c_R \Phi_0^{R,-}$ with $c_L c_R \ge 0, (c_L^2 + c_R^2 \neq 0)$.
Then, the general regular $S$-deformation such that $\tilde{V}$ vanishes is given by $S = - \Psi^{-1}\Psi^\prime$.
}
\\
{\bf Proof}.~~
From Proposition 3, $\Phi_0^{L,-}$ and $\Phi_0^{R,-}$ do not have a node, 
then $\Phi_0^{L,-} > 0$ and $\Phi_0^{R,-} > 0$.
Since it is clear that $\Psi$ does not have a node, 
we only need to show that the other solutions have a node.
The other solutions are given by $\tilde{\Psi} = a_1 \Phi_0^{L,-} + a_2 \Phi_0^{R,-}$ 
with $a_1 > 0, a_2 < 0$ (or $a_1 < 0, a_2 > 0$).
Since $\tilde{\Psi} \to a_2 \Phi_0^{R,-} < 0$ at $x \to -\infty$
and $\tilde{\Psi} \to a_1 \Phi_0^{L,-} > 0$ at $x \to \infty$,
$\tilde{\Psi}$ should have a node at some finite $x$.
\hfill$\Box$

\vspace{10pt}
Corollary 1. implies that the general regular $S$-deformation that satisfies Eq.~\eqref{sdefvzero} 
has a one-parameter degree of freedom.
Moreover, we can show the following:

\vspace{10pt}\hspace{-15pt}{\bf Proposition 4.}~~{\it 
If there exists no bound state with $E \le 0$,
defining $S_L = - \left(\Phi_0^{L,-}\right)^\prime/ \Phi_0^{L,-}, S_R = - \left(\Phi_0^{R,-}\right)^\prime/ \Phi_0^{R,-}$,
the general regular $S$-deformation such that $\tilde{V}$ vanishes 
can only take values between $S_L$ and $S_R$ at any point, 
and it can take all values between $S_L$ and $S_R$ at any point.
}
\\
{\bf Proof}.~~
From Corollary 1., the general regular $S$-deformation that satisfies Eq.\eqref{sdefvzero} 
can be written in the form
\begin{align}
S = S_L \sin^2\Theta 
+
 S_R \cos^2\Theta,
\end{align}
where $\sin\Theta, \cos\Theta$ are defined by
\begin{align}
\sin\Theta = \sqrt{\frac{c_L \Phi_0^{L,-}}{c_L \Phi_0^{L,-} + c_R \Phi_0^{R,-}}},
\quad
\cos\Theta = \sqrt{\frac{c_R \Phi_0^{R,-}}{c_L \Phi_0^{L,-} + c_R \Phi_0^{R,-}}},
\end{align}
with $0 \le \Theta \le \pi/2$.

At any point $x = a$, 
the functions $\Phi_0^{L,-}, \Phi_0^{R,-}$ are positive definite, 
and hence $\Theta$ can take all values in the range $0\le \Theta \le \pi/2$ by changing $c_L$ and $c_R$ 
with $c_L c_R \ge 0, (c_L^2 + c_R^2 \neq 0)$.
Since $S_R > S_L$ holds 
as shown in appendix~\ref{appendix:slsr}, 
the relations 
\begin{align}
S &= -(S_R - S_L)\sin^2\Theta + S_R \le S_R,
\\
S &= S_L + (S_R - S_L)\cos^2\Theta \ge S_L,
\end{align}
hold, and hence $S$ can take all values in $S_L \le S \le S_R$ at $x = a$ by changing $c_L$ and $c_R$.
Since $x = a$ is an arbitrary point, the statement in the present proposition holds.
\hfill$\Box$
\\

\begin{figure}[thbp]
\includegraphics[width=0.5\linewidth,clip]{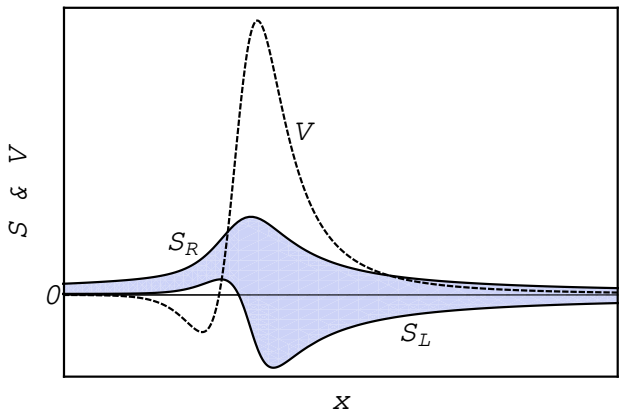}
 \caption{
The schematic figure for typical $S_L, S_R$ and $V$ (normalized by the horizon radius).
Two solid curves denote $S_L$ and $S_R$, and the dashed curve denotes $V$.
The shaded region corresponds to 
appropriate initial conditions for solving Eq.~\eqref{sdefvzero} 
to obtain regular $S$-deformations.
}
\label{fig:general_sdef}
\end{figure}

Proposition 4. implies that all initial conditions that give
regular $S$-deformations by solving Eq.~\eqref{sdefvzero} 
correspond to 
values between $S_L$ and $S_R$ at any point (see Fig.~\ref{fig:general_sdef}).
Thus, if we slightly change the initial condition in solving Eq.~\eqref{sdefvzero},
the solution still corresponds to a regular $S$-deformation.
This is the reason why we could obtain a regular $S$-deformation
by solving Eq.~\eqref{sdefvzero} numerically without any fine-tuning in~\cite{Kimura:2017uor}.
Typically, the potential $V$ becomes negative only near the horizon.
In this case, if the spacetime is stable, 
$S = 0$ at a point in the region far from the horizon becomes an appropriate initial condition 
to obtain a regular solution of Eq.~\eqref{sdefvzero} 
because $S_L$ is negative and $S_R$ is positive there like an example shown in Fig.~\ref{fig:general_sdef}.

\subsection{On the existence of zero mode}
Even if there exists a regular $S$-deformation such that $\tilde{V}$ vanishes,
there might still exist a zero mode, {\it i.e.,} $E = 0$ or $\omega = 0$ mode.\footnote{
If the zero mode asymptotes 
to a non-zero constant at either of $x \to \pm \infty$,
this is not normalizable.
However, this mode might be physically acceptable 
if $x \to \infty$ or $-\infty$ corresponds to the event horizon or cosmological horizon
which locates at the finite proper distance.
We note that this happens only when $V \to 0$ at $x \to \infty$ or $x \to - \infty$
and the boundary term in Eq.\eqref{sdef1} still vanishes since $S \to 0$ in this case.
}
We sometimes call $\omega = 0$ mode as a ``marginally stable mode'' because 
this is not an exponentially growing mode in time but a static perturbation.
However, since $t \Phi$ becomes also a solution of the wave equation \eqref{mastereqtr},
this implies the existence of a linearly growing mode in time.
So, it is important to discuss whether there exists $\omega = 0$ mode or not.

If there exists a zero mode $\Phi = \Phi_0$, from Eq.\eqref{sdef1}, $S$ should satisfy
\begin{align}
\frac{d\Phi_0}{dx} + S \Phi_0 = 0,
\label{zeromodewavefunc}
\end{align}
which
implies that there exists only a single $S$ if there exists a zero mode.
Thus, we can say the following:

\vspace{10pt}\hspace{-15pt}{\bf Proposition 5.}~~{\it 
If there exist two different regular functions $S = S_1, S_2$ 
such that $\tilde{V}$ vanishes, the zero mode does not exist.
}

\vspace{10pt}
This is a merit for the $S$-deformation method.
If we wish to show the non-existence of a zero mode by solving the Schr\"odinger equation directly,
we need to solve it from the decaying boundary condition, which is not always easy.

\section{Summary}
In this paper,
we showed the existence of 
a regular $S$-deformation function 
for a stable spacetime under some reasonable assumptions.
This implies that the $S$-deformation method 
can be used for showing the linear mode stability when the spacetime is stable,
and that the conjecture in the previous work~\cite{Kimura:2017uor} is correct.
We also showed that the general regular $S$-deformation such that $\tilde{V}$ vanishes 
contains a one-parameter degree of freedom.
This is the reason why 
we can find a regular $S$-deformation numerically without any fine-tuning~\cite{Kimura:2017uor}.

\section*{Acknowledgments}
We would like to thank Vitor Cardoso and Akihiro Ishibashi
for their useful discussion.
M.K. acknowledges financial support provided under 
the European Union's H2020 ERC Consolidator Grant 
``Matter and strong-field gravity: New frontiers in Einstein's theory''
grant agreement no. MaGRaTh-646597, 
and under the H2020-MSCA-RISE-2015 Grant No. StronGrHEP-690904.
T.T. acknowledges support in part by 
MEXT Grant-in-Aid for Scientific Research on Innovative Areas, Nos. 17H06357 and 17H06358, and by 
Grant-in-Aid for Scientific Research Nos. 26287044 and 15H02087.
M.K. also thanks Yukawa Institute for Theoretical Physics at Kyoto University 
and Kindai University for their hospitality.
We would like to thank the conference ``Black holes and strong gravity universe'',
where this work was initiated.

\appendix
\section{continuity of the position of zero as a function of $E$}
\label{appendix:continuityzeroofphi}

Even if $\Phi^{L,-}_E$ is not a continuous function of $E$,
we can show the position of the first zero of $\Phi^{L,-}_E$ 
is a continuous function of $E$ if it exists.
In this section, we denote $\Phi^{L,-}_E$ as $\Phi_E$ for simplicity.
When $\Phi_{E}$ has a zero, we denote the position by $\rho_E$,
{\it i.e.,} $\Phi_E|_{\rho_E} = 0$ and $\Phi_E > 0$ for $-\infty < x < \rho_E$.
Let us assume that $\rho_{E_0}$ exists for an energy $E_0 (<0)$.
For an energy $E < E_0$, 
from the same discussion in Lemma 2, we can say $\rho_E > \rho_{E_0}$ if it exists.

For $y > \rho_{E_0}$, we consider to solve the Schr\"odinger equation for $E \le E_0$ with the boundary condition
$\Phi|_y = 0, \Phi^\prime|_y =-1$. We denote this solution by $\phi_E$, and $\phi_E$ is
a continuous function of $E$ because
the boundary condition is given at the finite point $y$. 
Integrating the Wronskian conservation 
equation $(\phi_{E_0}^\prime \Phi_{E_0}- \Phi_{E_0}^\prime \phi_{E_0})^\prime = 0$ 
from $x = -\infty$, with sufficiently large $L>0$, to $x = \rho_{E_0}$, we obtain
$\Phi_{E_0}^\prime|_{\rho_{E_0}} \phi_{E_0}|_{\rho_{E_0}} \simeq 2 b_1 b_2 
\sqrt{V_- - {E_0}}$
where we used 
$\Phi_{E_0} \simeq b_1 e^{\sqrt{V_- - E_0}\, x}$ and 
$\phi_{E_0} \simeq b_2 e^{-\sqrt{V_- - E_0}\, x}$.\footnote{
Precisely speaking, 
$\Phi_{E_0} \simeq b_1 e^{\sqrt{V_- - E_0}\, x + f_+(x)}$ and 
$\phi_{E_0} \simeq b_2 e^{-\sqrt{V_- - E_0}\, x + f_-(x)}$, with $f_\pm = o(x)$, near $x \to -\infty$.
From the similar discussion in the proof of Lemma 4, we can say that 
$-\Phi_{E_0}^\prime/\Phi_{E_0}$ and $-\phi_{E_0}^\prime/\phi_{E_0}$ 
should be constants $- \sqrt{V_- - E_0}$ and $\sqrt{V_- - E_0}$ at $x \to -\infty$, respectively. 
Thus, $f_\pm^\prime = o(1)$ near $x \to -\infty$.
From the condition that $\phi_{E_0}^\prime \Phi_{E_0}- \Phi_{E_0}^\prime \phi_{E_0}$ 
is constant, $f_+ + f_- = o(1)$ should hold near $x \to -\infty$.
Hence, 
$\phi_{E_0}^\prime \Phi_{E_0}- \Phi_{E_0}^\prime \phi_{E_0} 
\simeq  -b_1 b_2 e^{f_+ + f_-}(2 \sqrt{V_m - E_0} + f_+^\prime - f_-^\prime)
\to -2b_1 b_2  \sqrt{V_m - E_0}$ at $x \to -\infty$,
and we can see that $o(x)$ terms do not affect the discussion.
}
Since $b_1 > 0$ and $\Phi_E^\prime|_{\rho_{E_0}} < 0$ from the assumption,
$b_2 \phi_{E_0}|_{\rho_{E_0}}$ should be negative.
Thus, $\phi_{E_0}$ has another zero in $x < \rho_{E_0} (< y)$,
and there exists $z$ such that $\phi_{E_0}|_z = 0$ and $\phi_{E_0} >0$ for $z < x < y$.
Since $\phi_E$ is a continuous function of $E$, $z$ is also a continuous function of $E$ if it exists.
Also, for an energy $E < V_{\rm min}$, where $V_{\rm min}$ is the minimum of $V$, 
$\phi_E$ does not have a zero except at $x = y$.
Thus, we can say that there exists $E_y~(< E_0)$ 
such that $z \to -\infty$ for $E = E_y$.

The asymptotic behavior near $x \to -\infty$ is
$\phi_{E} \simeq c_1 e^{-\sqrt{V_- - E}\, x} + c_2 e^{\sqrt{V_- - E}\, x}$.
If $0 < 1 - E/E_y \ll 1$, $z$ is in the asymptotic region.
So, from the condition $\phi_{E}|_{z} = 0$, 
we obtain 
$\phi_{E} \simeq c_2 (-e^{-\sqrt{V_- - E}(x - 2z)} + e^{\sqrt{V_- - E}x})$ when $0 < 1 - E/E_0 \ll 1$.
Therefore, in the limit $E \to E_y +0$, {\it i.e.,} $z \to -\infty$, the first term vanishes, 
and then $\phi_{E_y} \simeq c_1 e^{\sqrt{V_- - E_0}\, x} \to 0$ at $x \to -\infty$.
Thus, $\phi_{E_y} = c \Phi_{E_y}$ with a constant $c$.
This shows the existence of $\rho_E$ with $\infty > \rho_E > \rho_{E_0}$
since $y$ is an arbitrary point larger than $\rho_{E_0}$.

Also, since $\Phi_E$ does not have a zero except at $x \to -\infty$ for $E < V_{\rm min}$,
there should exist an energy $E_\infty$ such that $\rho_E \to \infty$ when $E \to E_\infty +0$.
Hence, $\rho_E$ is defined for $E_\infty < E \le E_0$. 
For any $E_1$ and $E_2$ with $E_0 \ge E_1 > E_2 > E_\infty$, there exist 
$\rho_{E_1}$ and $\rho_{E_2}$ with $\rho_{E_1} < \rho_{E_2}$.
Thus, $\rho_E$ is a continuous function of $E$ and
it monotonically increases as $E$ decreases.

\section{$S_R > S_L$}
\label{appendix:slsr}
Defining $S_L = - \left(\Phi_0^{L,-}\right)^\prime/ \Phi_0^{L,-}, S_R = - \left(\Phi_0^{R,-}\right)^\prime/ \Phi_0^{R,-}$, we can show $S_R > S_L$ if there exits no bound state with $E \le 0$.
Moreover, we can say stronger statement that $S_L$ is the minimum function near $x \to -\infty$
in the solutions of Eq.~\eqref{sdefvzero}.
Let us denote the general solution of the Schr\"odinger equation for $E = 0$ by $\Phi = z \Phi_0^{L,-}$,
then $z$ satisfies 
\begin{align}
z^{\prime \prime} + 2 \Big(\Phi_0^{L,-}\Big)^{-1}\Big(\Phi_0^{L,-}\Big)^{\prime} z^{\prime} =0.
\end{align}
Integrating this equation, $S = -\Phi^\prime/\Phi = S_L - z^\prime/z$ becomes
\begin{align}
S &= S_L + \Big(\Phi_0^{L,-}\Big)^{-2}\left(c_1 + \int_x^{c_2} dx \Big(\Phi_0^{L,-}\Big)^{-2}\right)^{-1}.
\end{align}
Note that $|c_1| \to \infty$ corresponds to the case $S = S_L$.
Since $\Phi_0^{L,-}$ is a positive constant or zero at $x \to -\infty$, 
the function $\int_x^{c_2} dx \left(\Phi_0^{L,-}\right)^{-2}$ is divergent at $x \to -\infty$.
Thus,
\begin{align}
S \simeq S_L + \Big(\Phi_0^{L,-}\Big)^{-2}\left(\int_x^{c_2} dx \Big(\Phi_0^{L,-}\Big)^{-2}\right)^{-1},
\end{align}
unless $|c_1| \to \infty$,
and the second term in the right hand side is positive.
This implies that all $S$ which satisfies Eq.~\eqref{sdefvzero} is larger than $S_L$ near $x \to -\infty$
except for the case $S = S_L$.

If there exits no bound state with $E \le 0$, 
both $S_L$ and $S_R$ are regular everywhere.
Also, $S_L$ cannot coincide with $S_R$ at a finite point because they are solutions
of Eq.~\eqref{sdefvzero} with different boundary conditions.
Thus, the relation $S_R > S_L$ holds everywhere in this case.

\end{document}